\documentclass{ws-p8-50x6-00_ladb}
\usepackage{here}
\usepackage{epsfig}
\usepackage{rotating}
\usepackage{graphicx}
\usepackage{amsmath}
\usepackage{amssymb}

\begin{document}

\title{CASTOR: A forward detector for the
       \protect\newline
       identification of Centauro and Strangelets
       \protect\newline
       in Nucleus--Nucleus Collisions at the LHC
       \footnote{
                 P\lowercase{resented at} XXVIII I\lowercase{nt.}
                 S\lowercase{ymp. on} M\lowercase{ultiparticle}
                 D\lowercase{ynamics,} 6-11 S\lowercase{ep.} 1998,
                 D\lowercase{elphi}.
                  \protect\newline\hspace*{4.5mm}
                 F\lowercase{urther information at:
    http://home.cern.ch/$^{\sim}$angelis/castor/}W\lowercase{elcome.html}
                 }
       }

\author{\vspace*{-4mm}
        A.L.S.~Angelis$^{\rm\lowercase{ a}}$,
        J.~Bartke$^{\rm\lowercase{ b}}$,
        M.Yu.~Bogolyubsky$^{\rm\lowercase{ c}}$,
        S.N.~Filippov$^{\rm\lowercase{ d}}$,
        \\
        E.~G\L{}adysz-Dziadu\'s$^{\rm\lowercase{ b}}$,
        Yu.V.~Kharlov$^{\rm\lowercase{ c}}$,
        A.B.~Kurepin$^{\rm\lowercase{ d}}$,
        \\
        A.I.~Maevskaya$^{\rm\lowercase{ d}}$,
        G.~Mavromanolakis$^{\rm\lowercase{ a}}$,
        A.D.~Panagiotou$^{\rm\lowercase{ a}}$,
        \\
        S.A.~Sadovsky$^{\rm\lowercase{ c}}$,
        P.~Stefanski$^{\rm\lowercase{ b}}$,
        Z.~W\L{}odarczyk$^{\rm\lowercase{ e}}$
        \vspace*{3pt}
       }

\address{$^{\rm\lowercase{ a}}$Nuclear and Particle Physics Division,
                               University of Athens, Hellas. \vspace*{-10pt}}
\address{$^{\rm\lowercase{ b}}$Institute of Nuclear Physics, Cracow, Poland.
                               \vspace*{-10pt}}
\address{$^{\rm\lowercase{ c}}$Institute for High Energy Physics, Protvino,
                               Russia. \vspace*{-10pt}}
\address{$^{\rm\lowercase{ d}}$Institute for Nuclear Research, Moscow, Russia.
                               \vspace*{-10pt}}
\address{$^{\rm\lowercase{ e}}$Institute of Physics, Pedagogical University, Kielce, Poland.}

\maketitle

\vspace*{-2mm}
\abstracts{
The physics motivation for a very forward detector to be employed in heavy
ion collisions at the CERN LHC is discussed.
A phenomenological model describing the formation and decay of a Centauro
fireball in nucleus-nucleus collisions is presented.
The CASTOR detector which is aimed to measure the hadronic and photonic
content of an interaction and to identify deeply penetrating objects in
the very forward, baryon-rich phase space 5.6~$\le~\eta~\le$~7.2 in an
event-by-event mode is described.
Results of simulations of the expected response of the calorimeter, and in
particular to the passage of strangelets, are presented.
          }
\vspace*{-8.5mm}

\section{Introduction}

The ALICE detector~\cite{Alice}, which is aimed at investigating
nucleus--nucleus collisions at the LHC, will be fully instrumented
for hadron and photon identification only in the limited
angular region around mid-rapidity, covering the pseudorapidity
interval $ | \eta | \le 1. $ An additional muon detector~\cite{Muons}
will be installed on one side and will cover the pseudorapidity
interval $ 2.5 \le \eta \le 4.0 $. A pre-shower photon multiplicity
detector~\cite{PMD} will also be installed on one side and will cover
the pseudorapidity interval $ 2.3 \le \eta \le 3.3 $.
Finally a small-aperture zero degree calorimeter is also foreseen,
at a distance of about 100 m from the interaction point, to provide
the centrality trigger. The very small acceptance and lack of longitudinal
and transverse segmentation in the present design of the zero degree
calorimeter make it suitable for this purpose only.

This constitutes only a small part of the total available phase
space which, at the design beam energy of 2.75 A TeV for $Pb$ ions
at the LHC, extends to $| \eta | = 8.7. $
Already at the early stages of the preparation of the ALICE proposal
some of us pointed out that there is interesting physics beyond
mid-rapidity~\cite{Note0,Gladysz1}. From these considerations evolved
the idea of a dedicated forward detector, CASTOR~\cite{DraftPro,Angelis1},
to provide physics information on hadrons and photons emitted in the
fragmentation region. At LHC $Pb$-ion energies this region of phase
space extends beyond $|\eta | =$ 5 or 6, depending on the effective
baryonic stopping power which is different for the various Monte-Carlo
models employed. In this region of extremely high baryon number density
one can expect to discover new phenomena related to the high baryochemical
potential, in particular the formation of Deconfined Quark Matter (DQM),
which could exist e.g. in the core of neutron stars, with characteristics
different from those expected in the much higher temperature baryon-free
region around mid-rapidity, 

The LHC, with an energy equivalent to $10^{17}$ eV for a moving proton
impinging on one at rest, will be the first accelerator to effectively
probe the highest cosmic ray energy domain. Cosmic ray experiments have
detected numerous most unusual events whose nature is still not understood. 
These events, observed in the projectile fragmentation rapidity region,
will be produced and studied at the LHC in controlled conditions.
Here we mention the ``Centauro'' events and the ``long-flying component''.
Centauros~\cite{Lates} exhibit relatively small multiplicity, complete
absence (or strong suppression) of the electromagnetic component and very
high $\langle p_{\rm T} \rangle$.
In addition, some hadron-rich events are accompanied by a strongly
penetrating component observed in the form of halo, strongly penetrating
clusters~\cite{Baradzei,Hasegawa} or long-living cascades, whose transition
curves exhibit a characteristic form with many maxima~\cite{Buja,Arisawa}.

\section{A model for the production of Centauro and Strangelets}

A model has been developed in which Centauros are considered to originate
from the hadronization of a DQM fireball of very high baryon density 
$\rm (\rho_b \gtrsim 2~fm^{-3})$ and baryochemical potential
$\rm (\mu_b >> m_n)$, produced in ultrarelativistic nucleus--nucleus
collisions in the upper atmosphere~\cite{Panagio1,Panagio2,Panagio3}.
In this model the DQM fireball initially consists of u, d quarks and gluons.
The very high baryochemical potential prohibits the creation of $\rm u\bar{u}$
and $\rm d\bar{d}$ quark pairs because of Pauli blocking of u and d quarks
and the factor exp~$\rm (-\mu_q/T)$ for $\rm \bar{u}$ and $\rm \bar{d}$
antiquarks, resulting in the fragmentation of gluons into s$\bar{\rm s}$
pairs predominantly. In the subsequent hadronization this leads to the
strong suppression of pions and hence of photons, but allows kaons to be
emitted, carrying away strange antiquarks, positive charge, entropy and
temperature. This process of strangeness distillation 
transforms the initial quark matter fireball into a slightly strange quark
matter state. In the subsequent decay and hadronization of this state
non-strange baryons and strangelets will be formed. Simulations show that
strangelets could be identified as the strongly penetrating particles
frequently seen accompanying hadron-rich cosmic ray
events~\cite{Gladysz1,Gladysz2}.
In this manner, both the basic characteristics of the Centauro events
(small multiplicities and extreme imbalance of hadronic to photonic
content) and the strongly penetrating component are naturally explained.
In table~\ref{tab:CRtoLHC} we compare characteristics of Centauro and 
strongly penetrating components (strangelets), either experimentally
observed or calculated within the context of the above model, for cosmic
ray interactions and for nucleus--nucleus interactions at the LHC.

\begin{table}[htb]
\begin{center}
\setlength{\tabcolsep}{1.5pc}
\vspace*{-3mm}
\caption{Average characteristic quantities of Centauro events and
Strangelets produced in Cosmic Rays and expected at the LHC.}
\label{tab:CRtoLHC}
\begin{tabular}{|c|c|c|}
\hline
  Centauro             &  Cosmic Rays             &  LHC                  \\
\hline
  Interaction          &  ``$Fe + N$''            & $ Pb + Pb $           \\
 $ \sqrt{s} $          & $ \gtrsim $ 6.76 TeV     &  5.5 TeV              \\
  Fireball mass        & $ \gtrsim $ 180  GeV     & $ \sim $ 500 GeV      \\
 $ y_{proj} $          & $ \geq $ 11              &   8.67                \\
 $ \gamma $            & $ \geq 10^4 $            & $ \simeq $ 300        \\
 $ \eta_{cent} $       &      9.9                 & $ \simeq $ 5.6        \\
 $ \Delta\eta_{cent} $ &      1                   & $ \simeq $  0.8       \\
 $ <p_T> $             &     1.75 GeV             &    1.75 GeV (*)       \\
  Life-time            & $ 10^{-9} $ s            & $ 10^{-9} $ s (*)     \\
  Decay prob.          & 10 \% (x $\geq$ 10 km)   & 1 \% (x $\leq$ 1 m)   \\
  Strangeness          &   14                     &  60 - 80              \\
 $ f_s $ (S/A)         & $ \simeq $ 0.2           & 0.30 - 0.45           \\
    Z/A                & $ \simeq $ 0.4           & $ \simeq 0.3 $        \\
 Event rate            & $ \gtrsim $ 1 \%         & $ \simeq $ 1000/ALICE-year \\
\hline
 ``Strangelet''        &  Cosmic Rays             &  LHC                  \\
\hline
  Mass                 & $ \simeq $ 7 - 15 GeV    & 10 - 80 GeV           \\
    Z                  & $ \lesssim $ 0           & $ \lesssim $ 0        \\
 $ f_s $               & $ \simeq $ 1             & $ \simeq $ 1          \\
 $ \eta_{str} $        & $ \eta_{cent} + 1.2 $    & $ \eta_{cent} + 1.6 $ \\
\hline
\end{tabular}
\end{center}
(*) assumed            
\vspace*{-5mm}
\end{table}

\section{The design of the CASTOR detector}

\begin{figure}[H]
\begin{center}
\parbox{0.48\hsize}{\epsfxsize=\hsize \epsfbox{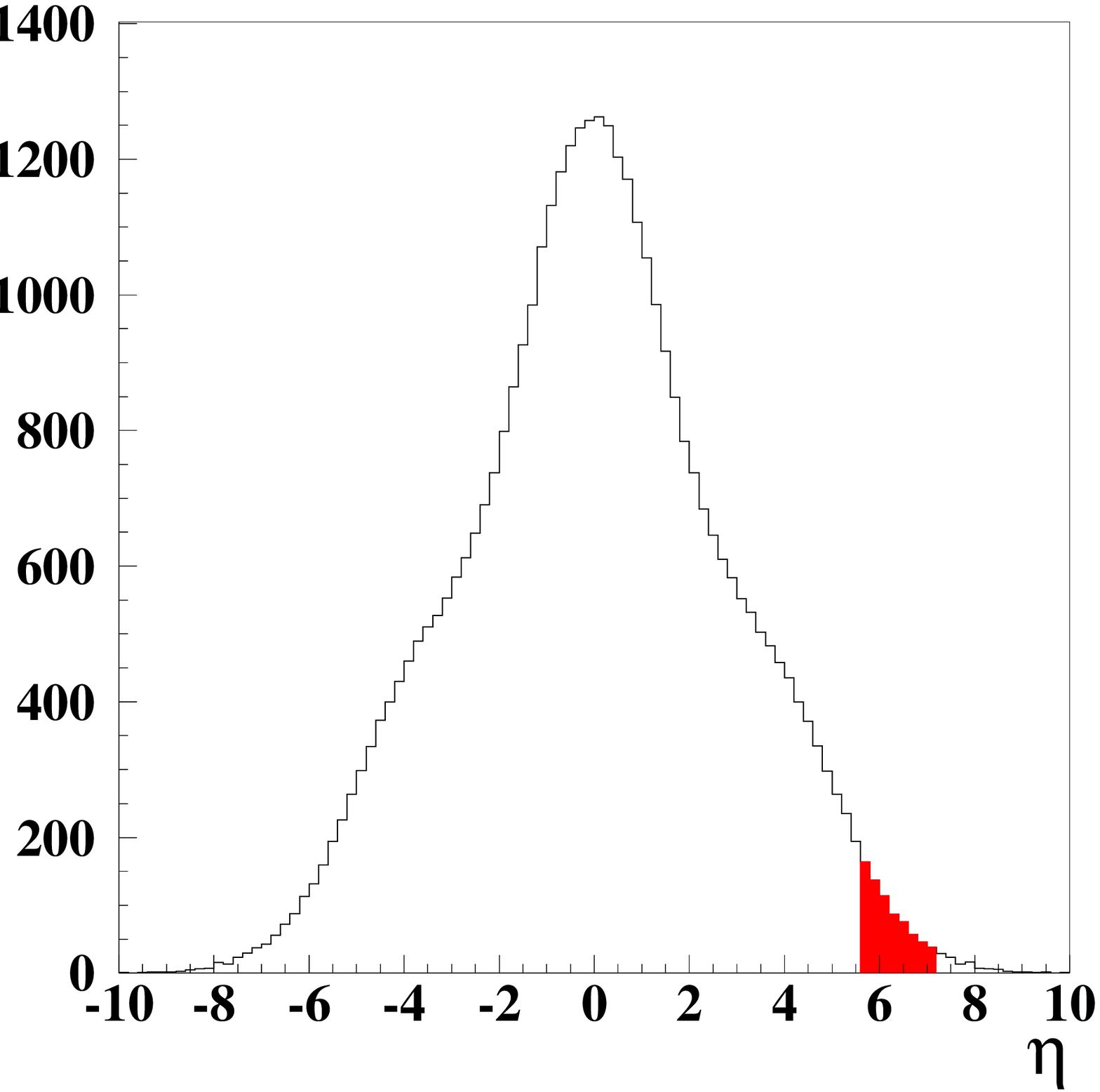}
                    \vspace*{-8mm}
                    \caption[]{Average photon multiplicity pseudorapidity
                               distribution obtained from 50 central $Pb+Pb$
                               HIJING events.}
                    \label{fig:Nf}
                   }
\hfill
\parbox{0.48\hsize}{\epsfxsize=\hsize \epsfbox{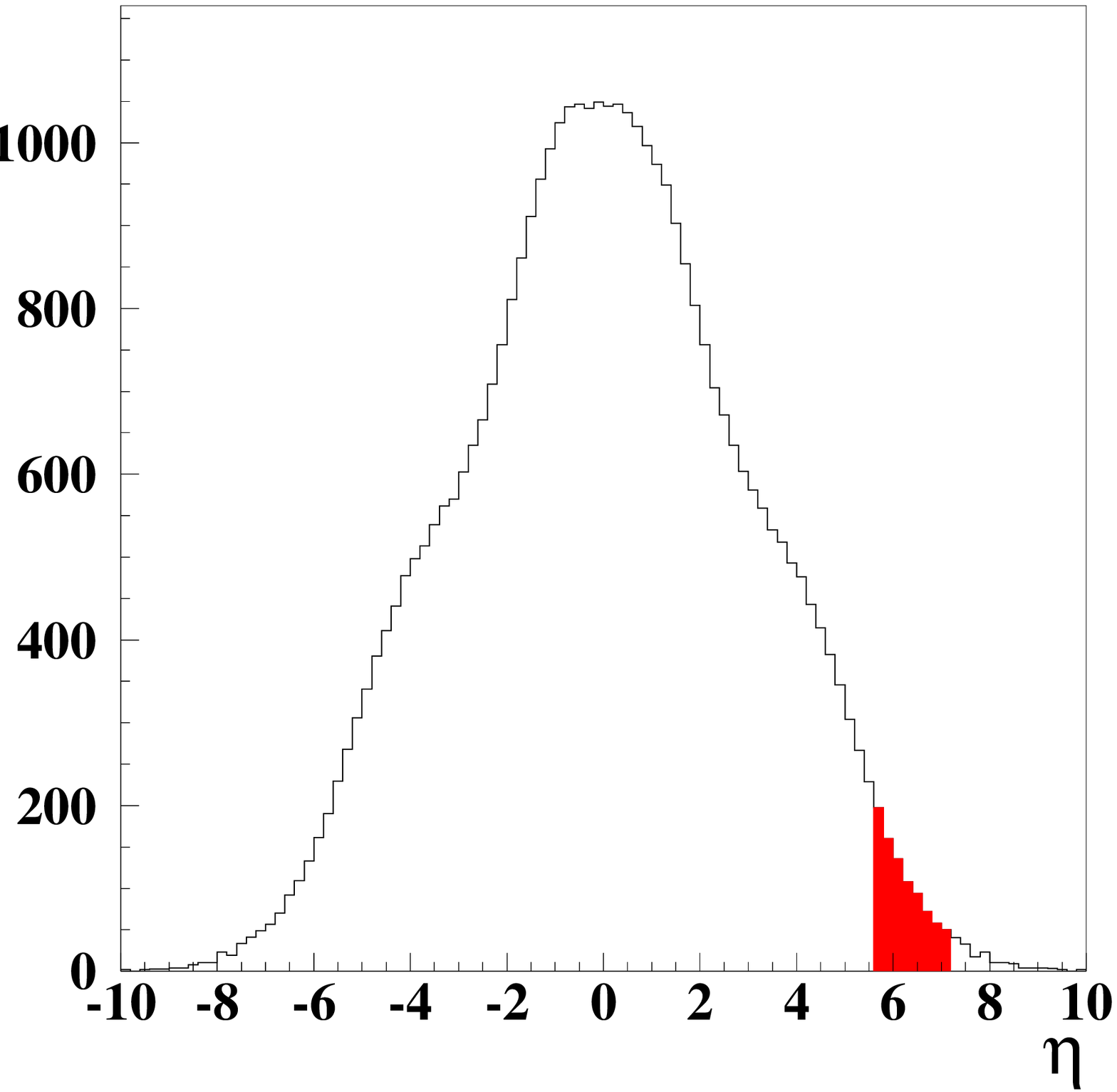}
                    \vspace*{-8mm}
                    \caption[]{Average charged particle multiplicity pseudorapidity
                               distribution obtained from 50 central $Pb+Pb$
                               HIJING events.}
                    \label{fig:Nch}
                   }
\end{center}
\end{figure}

\begin{figure}[H]
\begin{center}
\parbox{0.48\hsize}{\epsfxsize=\hsize \epsfbox{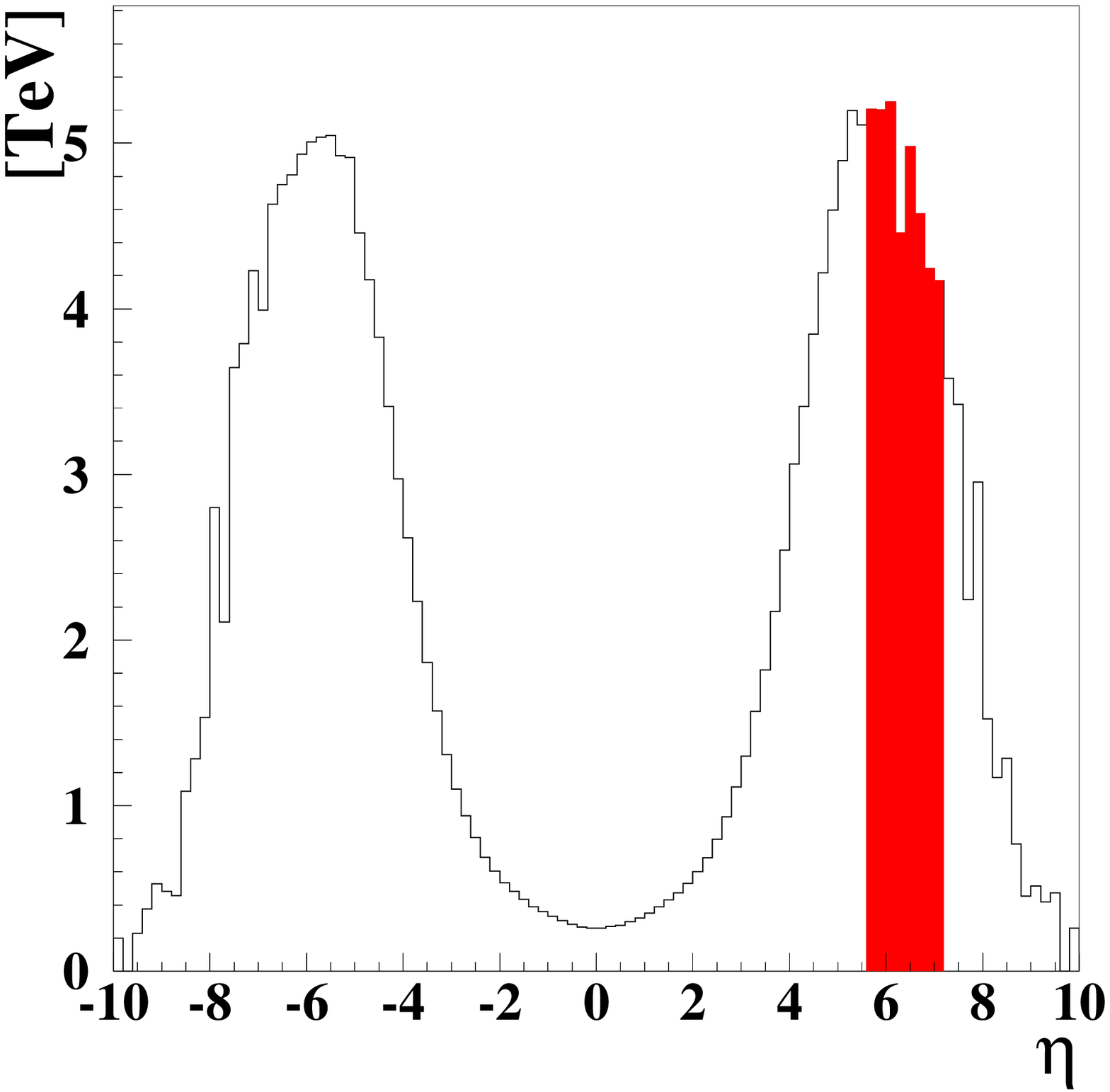}
                    \vspace*{-8mm}
                    \caption[]{Average electromagnetic energy pseudorapidity
                               distribution obtained from 50 central $Pb+Pb$
                               HIJING events.}
                    \label{fig:Eem}
                   }
\hfill
\parbox{0.48\hsize}{\epsfxsize=\hsize \epsfbox{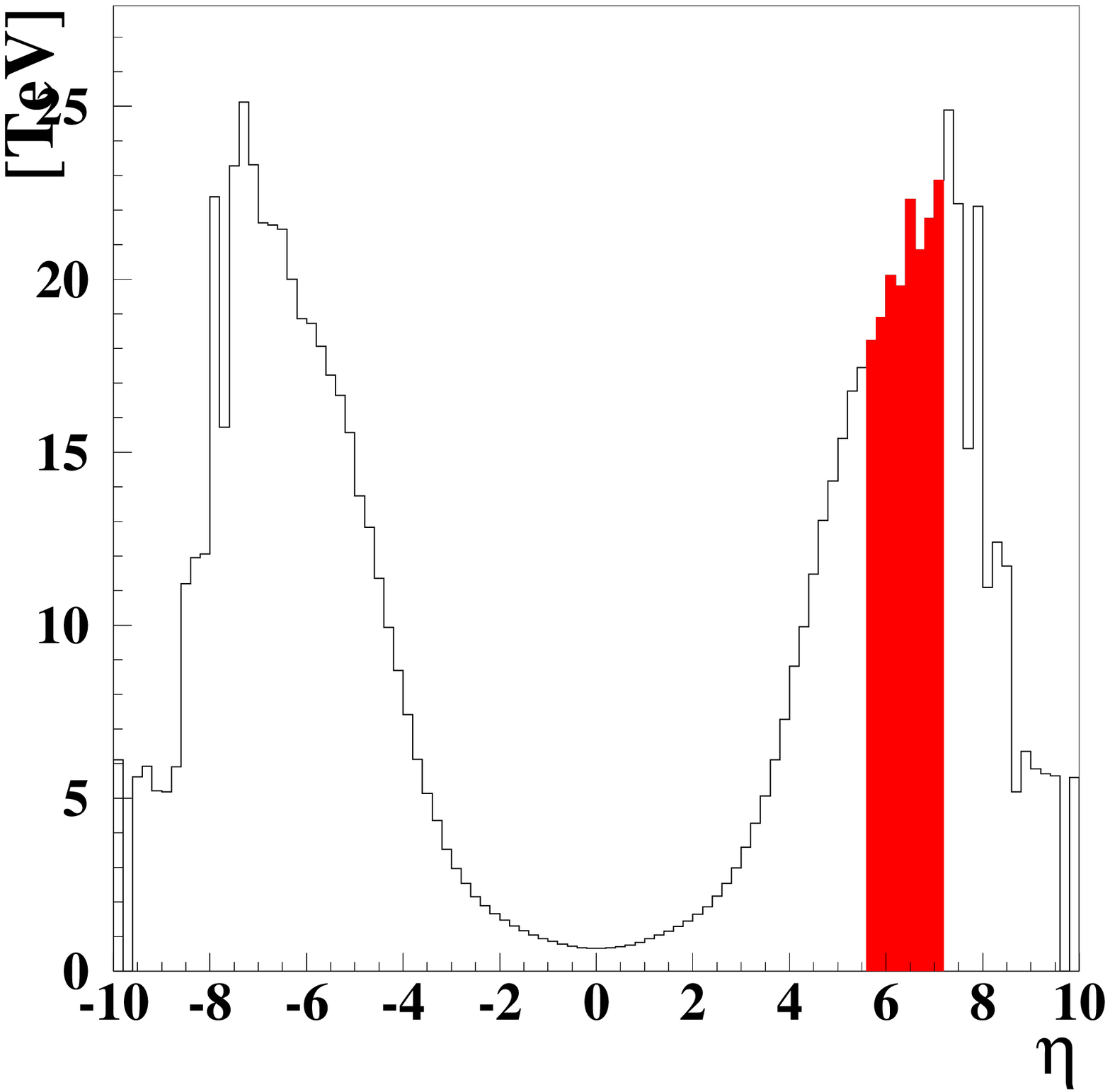}
                    \vspace*{-8mm}
                    \caption[]{Average hadronic energy pseudorapidity
                               distribution obtained from 50 central $Pb+Pb$
                               HIJING events.}
                    \label{fig:Ehad}
                   }
\end{center}
\end{figure}

\begin{figure}[H]
\begin{center}
\parbox{0.48\hsize}{\epsfxsize=\hsize \epsfbox{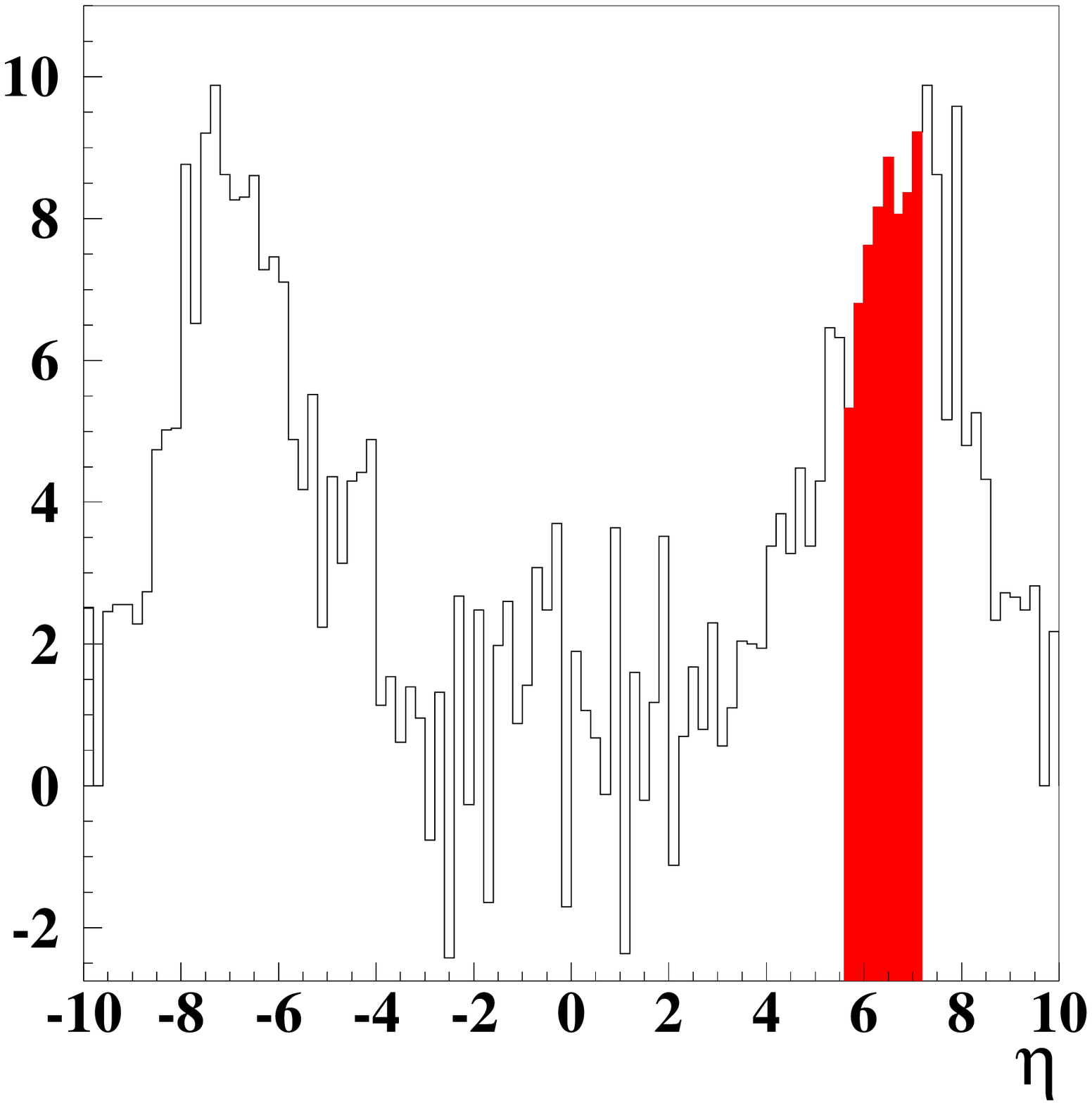}
                    \vspace*{-7mm}
                    \caption[]{Average net baryon number pseudorapidity
                               distribution obtained from 50 central $Pb+Pb$
                               HIJING events.}
                    \label{fig:Nb_hij}
                   }
\hfill
\parbox{0.48\hsize}{\epsfxsize=\hsize \epsfbox{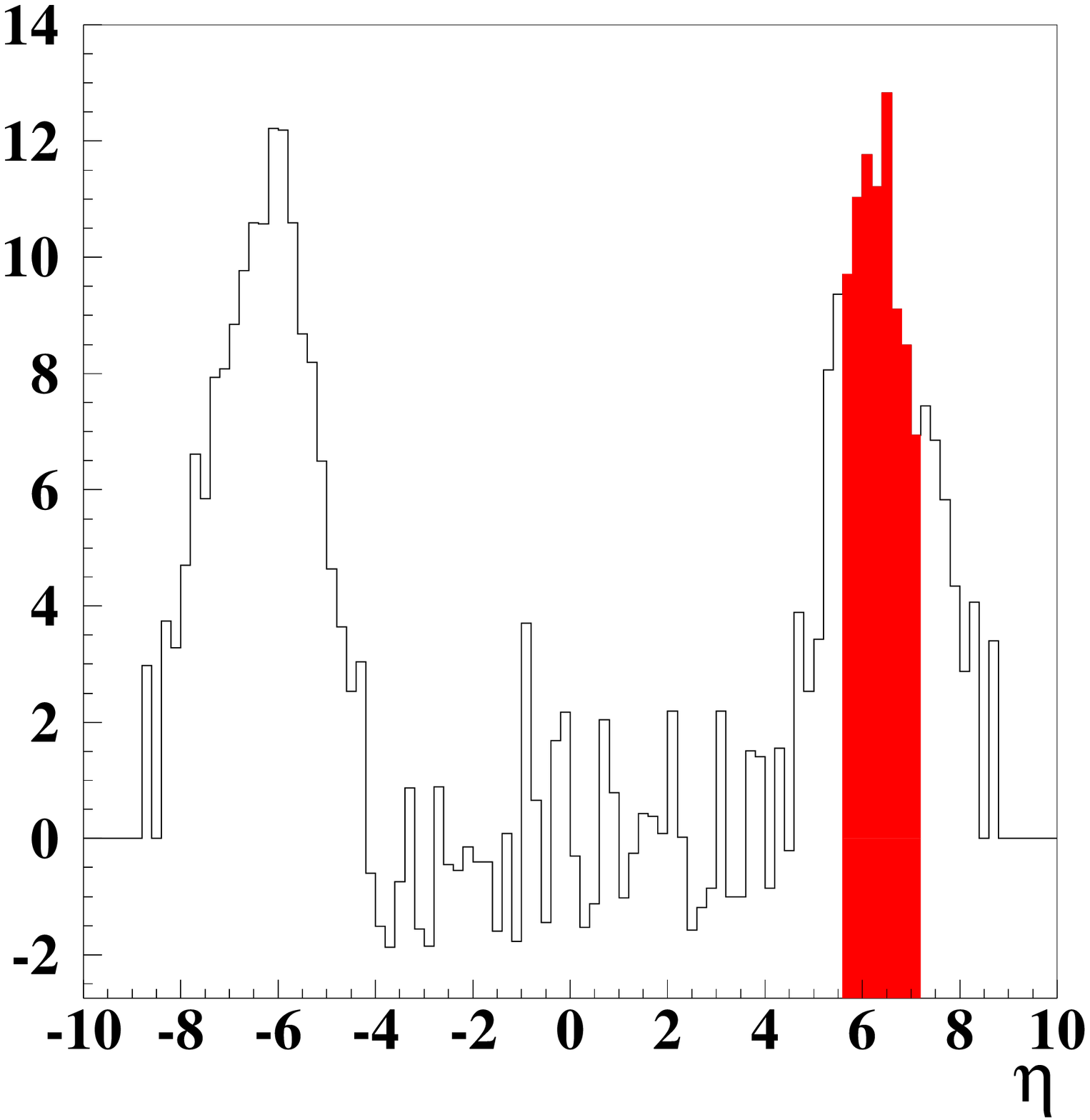}
                    \vspace*{-7mm}
                    \caption[]{Average net baryon number pdeudorapidity
                               distribution obtained from 47 central $Pb+Pb$
                               VENUS events.}
                    \label{fig:Nb_ven}
                   }
\end{center}
\end{figure}

With the above considerations in mind we have designed the CASTOR
(Centauro And STrange Object Research) detector, to be placed in the
fragmentation region. CASTOR will cover the pseudorapidity interval
$5.6 \le \eta \le 7.2$. Figures \ref{fig:Nf}, \ref{fig:Nch}, \ref{fig:Eem}
and \ref{fig:Ehad} depict the hadron and photon pseudorapidity distributions
as predicted by the HIJING Monte-Carlo generator for central $Pb+Pb$
collisions at the LHC. The upper plots show distributions of
multiplicity while the lower plots show distributions of energy flux. 
Figures \ref{fig:Nb_hij} and \ref{fig:Nb_ven} depict the baryon number
pseudorapidity distributions as predicted by the HIJING and VENUS
Monte-Carlo generators for central $Pb+Pb$ collisions at the LHC.
The acceptance of CASTOR is superimposed on each plot.
As can be seen from the plots, while CASTOR will receive a moderate
multiplicity, its position has been optimized to probe the maximum of
the baryon number density and energy flow and to identify any effects
connected with these conditions.

\vspace*{-6mm}

\vspace*{-14mm}

\begin{figure}[H]
\begin{center}
\vspace*{-2mm}
\includegraphics[width=9.5cm]{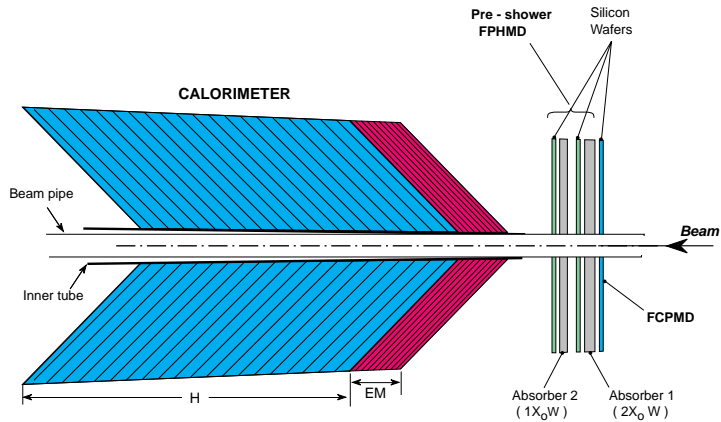}
\vspace*{-25mm}
\caption[]{Schematic representation of the CASTOR detector.} 
\label{fig:castor}
\end{center}
\end{figure}

\vspace*{-4mm}
A schematic view of the CASTOR detector~\cite{DraftPro,Angelis1} is
shown in figure~\ref{fig:castor}.
It is optimized to measure the hadronic and photonic content of an
interaction, both in energy and multiplicity, and to search for
strongly penetrating particles.
It will consist of a $Si$ pad charged particle multiplicity detector
followed by a $Si$ pad pre-shower photon multiplicity detector and of
a longitudinally segmented tungsten/quartz-fibre calorimeter with
electromagnetic and hadronic sections.

The multiplicity detectors have the form of annular discs of about 129 mm
outer and 26 mm inner radius, constructed in two half-rings to be positioned
around the beam pipe.

\begin{figure}[H]
\begin{center}
\begin{turn}{+90}
\includegraphics[width=4.6cm]{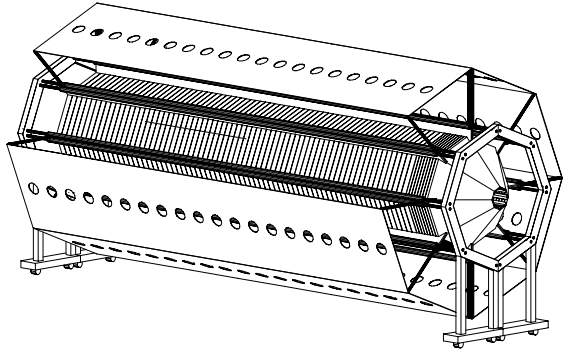}
\end{turn}
\vspace*{-2mm}
\caption[]{General view of the CASTOR calorimeter construction
           including support.} 
\label{fig:fcalo}
\end{center}
\end{figure}

The calorimeter is made of layers of active medium sandwiched between
tungsten absorber plates. The active medium consists of planes of silica
fibres and the signal is the Cherenkov light produced as they are
traversed by the charged particles in the shower.
The fibres are inclined at 45 degrees relative to the incoming particles
to maximize light output.
The calorimeter is azimuthally divided into 8 octants.
Each octant is longitudinally segmented into 80 layers, the first 8
($\simeq$ 14.7 X$_0$) comprising the electromagnetic section and the
remaining 72 ($\simeq$ 9.47 $\lambda_{\rm I}$) the hadronic section.
The light output from groups of 4 consecutive active layers is coupled
into the same light guide, giving a total of 20 readout channels along
each octant.
More detailed specifications are given in table~\ref{tab:CaloSpecs} and
a general view of the calorimeter, including its support, is shown in
figure~\ref{fig:fcalo}.
Mechanically the calorimeter is a structure built in two sections, left
and right, each consisting of four octants connected together at each
corner through bolted plates. The two sections and each of the octants
are self-supporting. It is envisaged to cut the edges of the absorber
plates in the azimuthal direction at an angle in such a way as to avoid
cracks between adjacent modules. The outer plates shown in
figure~\ref{fig:fcalo} (one omitted for clarity) constitute the support
for the light guides and photomultipliers.

\begin{table}[H]
\begin{center}
\vspace*{-3mm}
\caption{CASTOR calorimeter specifications.}
\label{tab:CaloSpecs}
\begin{tabular}{|c|c|c|}
\hline
                & Electromagnetic               & Hadronic                                        \\
\hline
Material        & Tungsten + Quartz Fibre       & Tungsten + Quartz Fibre                         \\
Dimensions      & $\rm \langle R_{in}\rangle = 26~mm,$  & $\rm \langle R_{in}\rangle = 27~mm,$  \\
                & $\rm \langle R_{out}\rangle = 129~mm$ & $\rm \langle R_{out}\rangle = 134~mm$ \\
Absorber Plates & Thickness = 5 mm              & Thickness = 10 mm                               \\
(at 45$^\circ$) & Eff. thickness = 7.07 mm      & Eff. thickness = 14.1 mm                        \\
No. Layers      &  8                            & 72                                              \\
Eff. length     & 56.6 mm $\simeq$ 14.7 X$_0$ $\simeq$ 0.53 $\lambda_{\rm I}$ 
                & 1018.1 mm $\simeq$ 9.47 $\lambda_{\rm I}$                                       \\
Quartz Fibre    & $\sim 0.45$ mm                & $\sim 0.45$ mm                                  \\
No. QF planes   &  2 per sampling               &  4 per sampling                                 \\
Sampling        & $\simeq$ 1.84 X$_0$           & $\simeq$ 0.13 $\lambda_{\rm I}$                 \\
Reading         & Coupling of 4 samplings       & Coupling of 4 samplings                         \\
No. Readings    &  2                            & 18                                              \\
No. Channels    & $ 2 \times 8 = 16$            & $18 \times 8 = 144$                             \\
QF/W vol.       & 10\%                          & 10\%                                            \\
\hline
\end{tabular}
\end{center}
\end{table} 

\vspace*{-9mm}

\section{Simulation of the CASTOR calorimeter performance}

\begin{figure}[H]
\begin{center}
\parbox{0.48\hsize}{\epsfxsize=\hsize \epsfbox{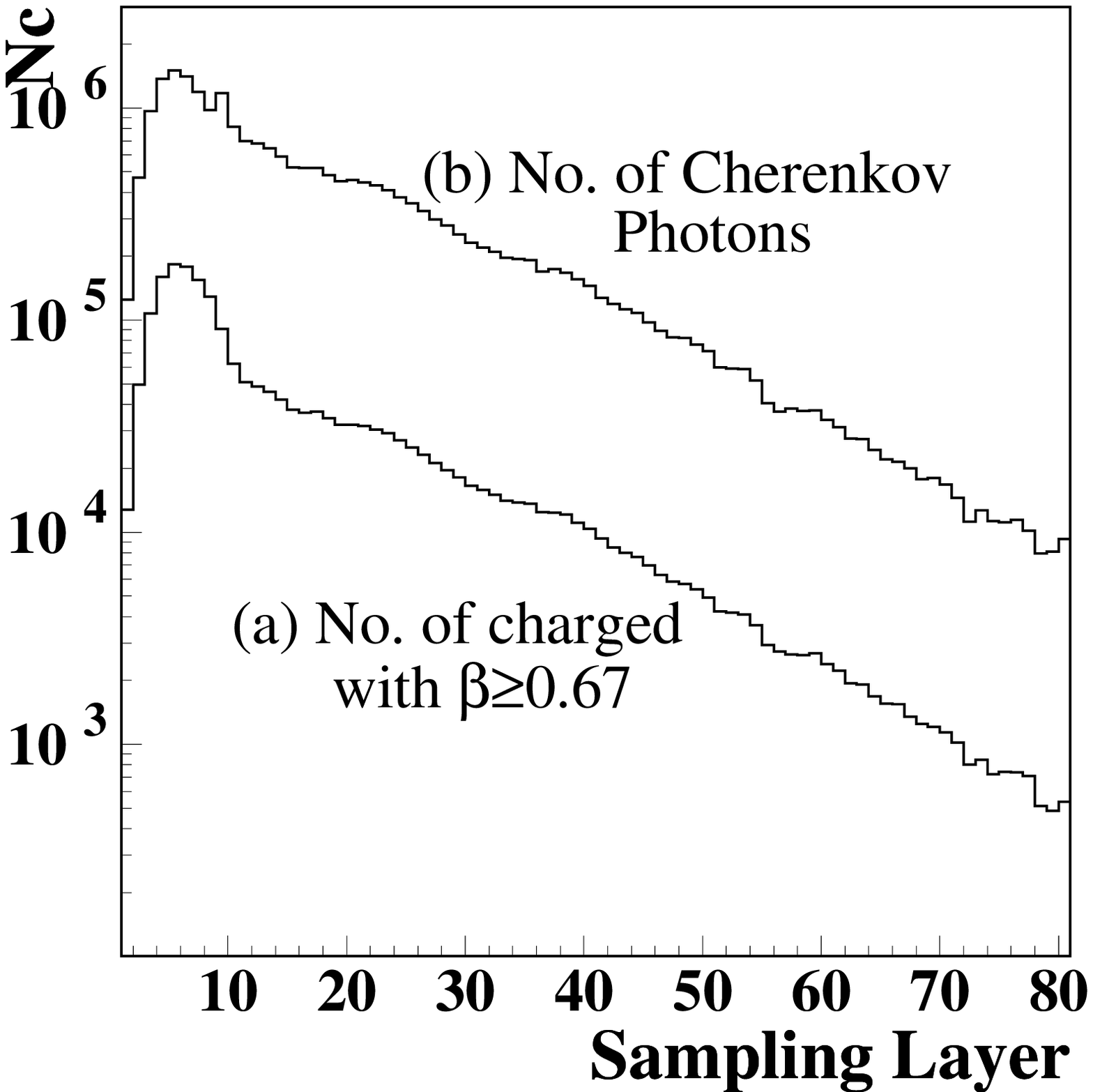}
                    \vspace*{-7mm}
                    \caption[]{Response of the CASTOR calorimeter:
                               (a) Number of charged particles above Cherenkov
                                   threshold at each active layer,
                               (b) Number of Cherenkov photons produced and
                                   propagated inside the fibres at each active
                                   layer.}
                    \label{fig:simul1}
                   }
\hfill
\parbox{0.48\hsize}{\epsfxsize=\hsize \vspace*{-5mm} \epsfbox{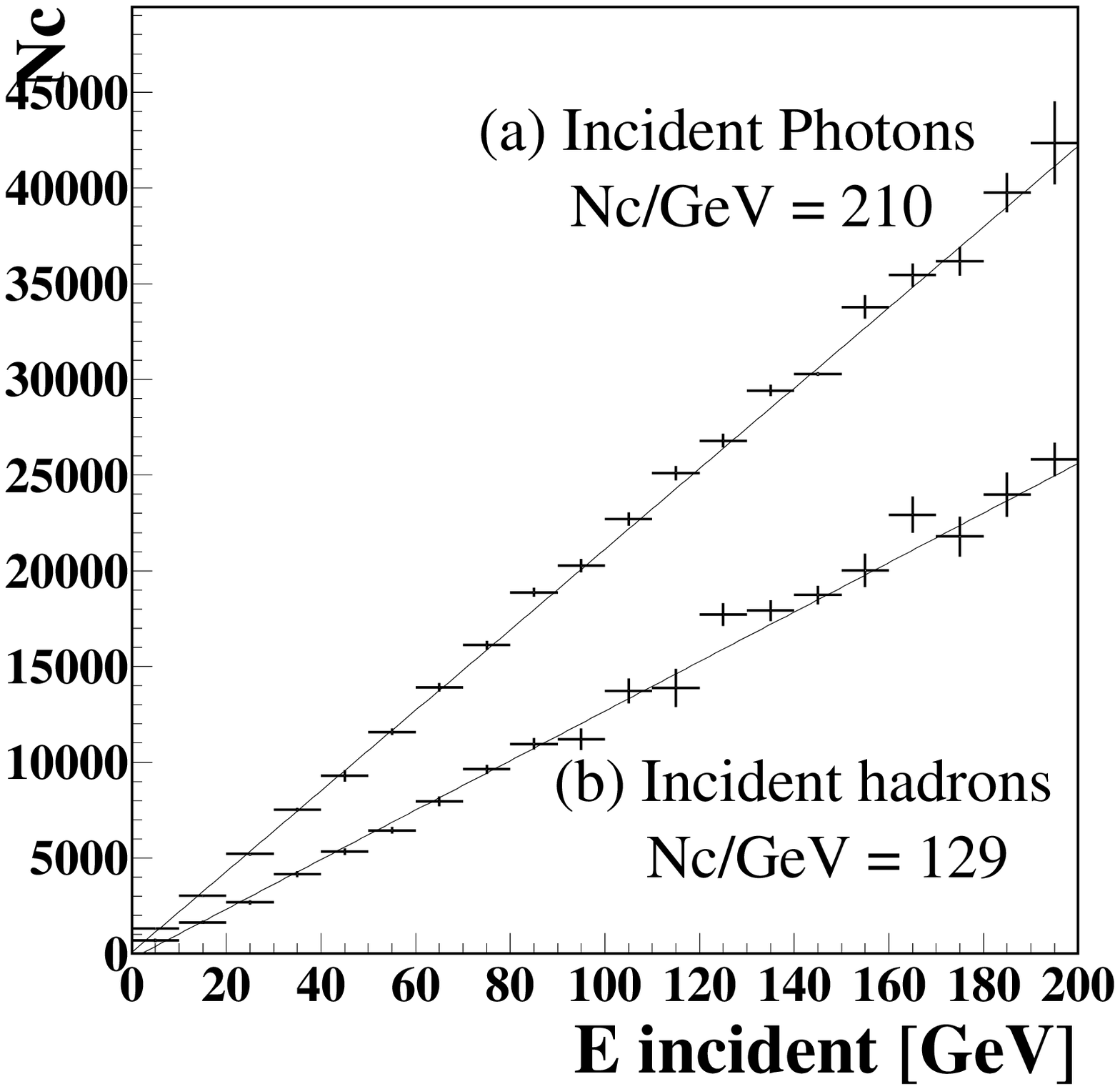}
                    \vspace*{-6mm}
                    \caption[]{Total number of Cherenkov photons produced, 
                               captured and propagated inside the fibres,
                               vs. incident particle energy:
                         (a) For incident photons, (b) For incident hadrons.}
                    \label{fig:simul2}
                   }
\end{center}
\end{figure}

We have made detailed GEANT simulations of the performance of the
C~A~S~T~O~R calorimeter.
Figure~\ref{fig:simul1} shows its response to one central $Pb+Pb$ HIJING event:
figure~\ref{fig:simul1}a shows the number of charged particles (essentially
$e^+,e^-$) above Cherenkov threshold in the showers at each active layer,
while figure~\ref{fig:simul1}b shows the corresponding number of Cherenkov
photons which are produced, captured and propagated inside the fibres.

\vspace*{-5mm}
\vspace*{-4mm}

Figure~\ref{fig:simul2} shows the total number of Cherenkov photons produced,
captured and propagated inside the fibres, as a function of the incident
particle energy, for incident photons and hadrons from one central $Pb+Pb$
HIJING event. About 210 Cherenkov photons per GeV are obtained for incident
photons and 129 Cherenkov photons per GeV for incident hadrons.

\begin{figure}[H]
\begin{center}
\includegraphics[width=12cm]{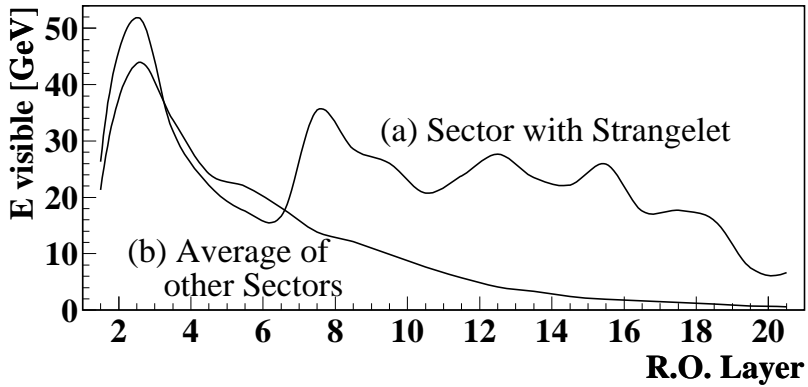}
\vspace*{-63mm}
\caption[]{Energy deposition in the readout layers (couplings of 4
           consecutive sampling layers) of the CASTOR calorimeter:
           (a) In the octant containing the Strangelet,
           (b) Average of the other octants.}
\label{fig:Slet}
\end{center}
\end{figure}

In addition we have simulated the interaction of a Strangelet with the
calorimeter material, using the simplified picture described
in~\cite{Gladysz2,Angelis2}.
As an example figure~\ref{fig:Slet} shows the response of the calorimeter
to one central $Pb+Pb$ HIJING event, which contains a Strangelet of 
$\rm A_{str}$=20, $\rm E_{str}$=20 TeV and quark chemical potential
$\rm \mu_{str}$=600 MeV (energy conservation has been taken into account).
Figure~\ref{fig:Slet}a shows the energy deposition along the octant
containing the Strangelet,
while figure~\ref{fig:Slet}b shows the average of the energy deposition 
along the other seven octants.

The study of such simulated events shows that the signal from an octant
containing a Strangelet is larger than the average of the others, while
its transition curve displays long penetration and many maxima structure,
such as observed in cosmic ray events.

\vspace*{-12mm}
\vspace*{-8mm}

\section{Conclusions}

We have designed a detector system well suited to probe the very forward
region in $Pb+Pb$ collisions at the LHC, where very large baryon number
density and energy flow occur.
Our detector will identify any effects connected with these conditions.
It has been particularly optimised to search for signatures of Centauro
and for long penetrating objects.
We have developed a model which explains Centauro production in cosmic rays
and makes predictions for $Pb+Pb$ collisions at the LHC.
Our model naturally incorporates the possibility of Strangelet formation.
We have simulated the passage of Strangelets through the CASTOR calorimeter
and we find long penetration and many-maxima structures similar to those
observed in cosmic ray events.

\section*{Acknowledgements}
This work has been partly supported by the Hellenic General Secretariat
for Research and Technology $ \rm \Pi ENE \Delta $ 1361.1674/31-1/95,
the Polish State Committee for Scientific Research grants 2P03B 121 12
and SPUB P03/016/97,
and the Russian Foundation for Fundamental Research grant 96-02-18306.

\end{document}